# An Algebraic Theory of Complexity for Discrete Optimisation[*]

David A. Cohen
Department of Computer Science
Royal Holloway, University of London, UK
d.cohen@rhul.ac.uk

Martin C. Cooper
IRIT
University of Toulouse III, France
cooper@irit.fr

Páidí Creed
School of Mathematical Sciences
Queen Mary, University of London, UK
p.creed@qmul.ac.uk

Peter G. Jeavons
Department of Computer Science
University of Oxford, UK
peter.jeavons@cs.ox.ac.uk

Stanislav Živný
University College
University of Oxford, UK
standa.zivny@cs.ox.ac.uk

**Abstract**

Discrete optimisation problems arise in many different areas and are studied under many different names. In many such problems the quantity to be optimised can be expressed as a sum of functions of a restricted form. Here we present a unifying theory of complexity for problems of this kind. We show that the complexity of a finite-domain discrete optimisation problem is determined by certain algebraic properties of the objective function, which we call *weighted polymorphisms*. We define a Galois connection between sets of rational-valued functions and sets of weighted polymorphisms and show how the closed sets of this Galois connection can be characterised.

These results provide a new approach to studying the complexity of discrete optimisation. We use this approach to identify certain maximal tractable subproblems of the general problem, and hence derive a complete classification of complexity for the Boolean case.

## 1 Introduction

Discrete optimisation problems arise in many different areas and are studied under many different names, including Min-Sum Problems, Gibbs energy minimisation, Markov Random Fields, Conditional Random Fields, 0/1 integer programming, pseudo-Boolean function minimisation, constraint optimisation and valued constraint satisfaction [31, 10, 40, 39, 17, 16, 7, 6].

[*]Part of this work (by D. A. Cohen, M. C. Cooper, and P. G. Jeavons) appeared in *Proceedings of the 12th International Conference on Principles and Practice of Constraint Programming (CP)*, pp. 107–121, 2006. Part of this work (by D. A. Cohen, P. Creed, P. G. Jeavons, and S. Živný) appeared in *Proceedings of the 36th International Symposium on Mathematical Foundations of Computer Science (MFCS)*, pp. 231-242, 2011. Part of this work (by P. Creed and S. Živný) appeared in *Proceedings of the 17th International Conference on Principles and Practice of Constraint Programming (CP)*, pp. 210–224, 2011. Part of this work appeared in Stanislav Živný's doctoral thesis at the University of Oxford, 2009. This research was supported by EPSRC grant EP/F01161X/1. Stanislav Živný is supported by a Junior Research Fellowship at University College, Oxford.

Here we adopt a very general framework where each problem instance is specified by a set of variables, a set of possible values for those variables, and a set of constraints. Each combination of values allowed by each constraint has an associated *cost*, and the goal is to find an assignment with minimal total cost. This simple abstract mathematical framework can be used to express discrete optimisation problems arising in a wide variety of fields, including operational research (scheduling, resource utilisation, transportation), computer vision (region segmentation, object recognition, image enhancement), automated reasoning (Max SAT, Min ONES), graph theory (Min-Cut, Maximum Independent Set), and many others.

In the special case when all costs are zero, the problem we are studying collapses to the standard constraint satisfaction problem (CSP). The general CSP is NP-hard, and so is unlikely to have a polynomial-time algorithm. However, there has been much success in finding tractable fragments of the CSP by restricting the types of relation allowed in the constraints. A set of allowed relations has been called a *constraint language* [27]. For some constraint languages the associated constraint satisfaction problems with constraints chosen from that language are solvable in polynomial-time, whilst for other constraint languages this class of problems is NP-hard [28, 27, 22]; these two cases are referred to as *tractable languages* and *NP-hard languages*, respectively. Dichotomy theorems, which classify each possible constraint language as either tractable or NP-hard, have been established for languages over 2-element domains [34], 3-element domains [13], for conservative languages [15, 1], and maximal languages [11, 12].

The more general framework we consider here, which allows non-zero costs, is also NP-hard, but again we can try to identify tractable fragments by restricting the types of allowed constraints. Each type of constraint is specified by a rational-valued function defined on a set of tuples, which specifies the cost associated with each allowed tuple of values. Such a function is called a *weighted relation*, and a set of weighted relations will be called a *valued constraint language* [16]. Much less is known about the complexity of the optimisation problems associated with different valued constraint languages, although some results have been obtained for certain special cases. In particular, a complete characterisation of complexity has been obtained for valued constraint languages over a 2-element domain [16]. This result generalises a number of earlier results for particular optimisation problems such as MAX-SAT [18] and MIN-ONES [19]. A complete classification has also been obtained for valued constraint languages containing all unary $\{0,1\}$-valued weighted relations (such languages are called *conservative*) [30]. This result generalises a number of earlier results such as GRAPH MIN-COST-HOM [25] and DIGRAPH MIN-COST-HOM [37].

One class of weighted relations that has been extensively studied and shown to be tractable is the class of *submodular* functions [19, 16, 29, 21, 30, 41].

In the classical CSP framework it has been shown that the complexity of any constraint language over any finite domain is determined by certain algebraic properties known as *polymorphisms* [28, 27]. This result has reduced the problem of the identification of tractable constraint languages to that of the identification of suitable sets of polymorphisms. In other words, it has been shown to be enough to study just those constraint languages which are characterised by having a given set of polymorphisms. Using this algebraic approach, considerable progress has now been made towards a complete characterisation of the complexity of constraint languages over finite domains of arbitrary size [22, 14, 4, 2, 3, 5].

In this paper, we introduce a new algebraic construct which we call a *weighted polymorphism*. We are able to show that the weighted polymorphisms of a valued constraint language are sufficient to determine the complexity of that language. In addition, we are able to define a Galois connection between valued constraint languages and sets of weighted polymorphisms, and characterise the closed sets on both sides.

The Galois connection we establish here can be used in the search for tractable valued constraint languages in a very similar way to the use of polymorphisms in the search for tractable constraint languages in the classical CSP. First, we need only consider valued constraint languages characterised by weighted polymorphisms. This greatly simplifies the search for a characterisation of all tractable valued constraint languages. Second, we will show below that any tractable valued constraint language must have an associated non-trivial weighted polymorphism. Hence the results of this paper provide a powerful new set of tools for analysing the complexity of finite-domain discrete optimisation problems. In fact, a recent result on the power of linear programming for valued constraint languages builds on weighted polymorphisms [38].

The structure of the paper is as follows. In Section 2 we introduce the general framework of the Valued Constraint Satisfaction Problem (VCSP) and define the notion of expressibility. In Section 3 we focus on the classical Constraint Satisfaction Problem and show how it fits in the VCSP framework as a special case. We briefly recall the notion of polymorphism, and the Galois connection that has been so fruitful in the study of the complexity of the classical Constraint Satisfaction Problem. In Sections 4 and 5 we introduce *weighted relational clones* (valued constraint languages closed under expressibility and certain other operations) and the corresponding closed sets of weighted polymorphisms, which we call *weighted clones*. We then state our main result: weighted relational clones are in 1-to-1 correspondence with weighted clones. In Section 6 we give proofs of the theorems establishing this Galois connection. In Section 7 we use the algebraic theory to establish necessary conditions that must be satisfied by any tractable valued constraint language. Using these results, we obtain a complete classification for the Boolean case in Section 8. Finally, in Section 9, we state some conclusions and outline directions for future work.

## 2 The Valued Constraint Satisfaction Problem

For any set $D$, the set of tuples of length $r$ over $D$ is denoted by $D^r$, and a subset of $D^r$ is called a *relation* over $D$ of arity $r$. A (partial) function $\varrho$ from $D^r$ to $\mathbb{Q}$ associates a rational[1] weight with each of the tuples in some subset of $D^r$, and so will be called a **weighted relation** on $D$ of *arity $r$*.

The idea of a weighted relation is very general, and can be used to define a wide variety of discrete optimisation problems. The general framework we shall use for such problems is defined as follows.

**Definition 2.1.** *An instance of the* **valued constraint satisfaction problem**, *(VCSP), is a triple $\mathcal{P} = \langle V, D, C \rangle$ where: $V$ is a finite set of* **variables**; *$D$ is a finite set of possible* **values**; *$C$ is a finite multi-set of* **constraints**. *Each element of $C$ is a pair $c = \langle \sigma, \varrho \rangle$ where $\sigma$ is a tuple of variables called the* **scope** *of c, and $\varrho$ is a weighted relation on $D$ of arity $|\sigma|$.*

*An* **assignment** *for $\mathcal{P}$ is a mapping $s : V \to D$. The* **cost** *of an assignment $s$, denoted $Cost_\mathcal{P}(s)$, is given by the sum of the weights assigned to the restrictions of $s$ onto each constraint scope, that is,*

$$Cost_\mathcal{P}(s) \stackrel{\text{def}}{=} \sum_{\langle \langle v_1, v_2, \ldots, v_m \rangle, \varrho \rangle \in C} \varrho(s(v_1), s(v_2), \ldots, s(v_m)).$$

*If $\varrho(s(v_1), s(v_2), \ldots, s(v_m))$ is undefined for some $\langle \langle v_1, v_2, \ldots, v_m \rangle, \varrho \rangle \in C$, then the assignment $s$ is said to be* infeasible *and $Cost_\mathcal{P}(s)$ is undefined.*

*A* **solution** *to $\mathcal{P}$ is a feasible assignment with minimal cost.*

**Example 2.2** (MIN-CUT). *In graph theory a* **cut** *of a graph is a partition of the vertices into two disjoint sets. The size of a cut is the number of edges of the graph that intersect both sides of this partition. The MIN-CUT problem for a graph is to find a cut with the smallest possible size. This problem can be solved in polynomial-time [32].*

*The MIN-CUT problem for the graph $(V, E)$ can be expressed as an instance $\langle V, \{0, 1\}, C \rangle$ of VCSP by setting $C = \{\langle e, \varrho_= \rangle \mid e \in E\}$, where $\varrho_=$ is the binary weighted relation on the set $\{0, 1\}$ defined by:*

$$\varrho_=(x, y) \stackrel{\text{def}}{=} \begin{cases} 0 & \text{if } x = y \\ 1 & \text{otherwise} \end{cases}.$$

---

[1] To avoid representational issues, we restrict ourselves to rational rather than real-valued weights. The resulting framework is sufficiently general to encode very many standard optimisation problems; for examples, see [16].

*Any assignment for this VCSP instance partitions the elements of $V$ into those assigned the value 0 and those assigned the value 1. The cost of the assignment is equal to the size of the corresponding cut.*

**Example 2.3** (MAX-CUT). *The MAX-CUT problem for a graph is to find a cut with the largest possible size. This problem is NP-hard [23].*

*The MAX-CUT problem for the graph $(V, E)$ can be expressed as an instance $\langle V, \{0, 1\}, C \rangle$ of VCSP by setting $C = \{\langle e, \varrho_{\neq} \rangle \mid e \in E\}$, where $\varrho_{\neq}$ is the binary weighted relation on the set $\{0, 1\}$ defined by:*

$$\varrho_{\neq}(x, y) \stackrel{\text{def}}{=} \begin{cases} 0 & \text{if } x \neq y \\ 1 & \text{otherwise} \end{cases}.$$

**Example 2.4** (DIGRAPH MIN-COST-HOM). *Given two directed graphs (digraphs) $G = (V_G, E_G)$ and $H = (V_H, E_H)$, a homomorphism from $G$ to $H$ is a mapping $f : V_G \to V_H$ that preserves edges, that is, $(u, v) \in E_G$ implies $(f(u), f(v)) \in E_H$. Assume that for any $u \in V_G$ and $v \in V_H$ a rational cost $c_u(v)$ is given. The cost of a homomorphism $f$ from $G$ to $H$ is then defined to be $\sum_{u \in V_G} c_u(f(u))$. The DIGRAPH MIN-COST-HOM problem is to find a homomorphism from $G$ to $H$ of minimum cost [25, 37].*

*Given a fixed digraph $H = (V_H, E_H)$, we denote by $\varrho_H$ the binary weighted relation on the set $V_H$ defined by:*

$$\varrho_H(x, y) \stackrel{\text{def}}{=} \begin{cases} 0 & \text{if } (x, y) \in E_H \\ \infty & \text{otherwise} \end{cases}.$$

*The DIGRAPH MIN-COST-HOM problem for input graph $G = (V_G, E_G)$ and fixed target graph $H = (V_H, E_H)$ can be expressed as an instance $\langle V_G, V_H, C \rangle$ of VCSP by setting $C = \{\langle e, \varrho_H \rangle \mid e \in E_G\} \cup \{\langle u, c_u \rangle \mid u \in V_G\}$.*

A **valued constraint language** is any set $\Gamma$ of weighted relations on some fixed set $D$. We define VCSP($\Gamma$) to be the set of all VCSP instances in which all weighted relations in all constraints belong to $\Gamma$.

Example 2.2 shows that VCSP($\{\varrho_=\}$) includes the MIN-CUT problem, and Example 2.3 shows that VCSP($\{\varrho_{\neq}\}$) includes the MAX-CUT problem. In fact it is easy to show that VCSP($\{\varrho_=\}$) corresponds to the MIN-CUT problem in the sense that not only does VCSP($\{\varrho_=\}$) include the MIN-CUT problem, but also any instance from VCSP($\{\varrho_=\}$) can be reduced to an instance of the MIN-CUT problem. Similarly, VCSP($\{\varrho_{\neq}\}$) corresponds to the MAX-CUT problem. The problem VCSP($\{\varrho_=, \varrho_{\neq}\}$) includes both MIN-CUT and MAX-CUT, as well as many other problems which can be expressed using these two types of constraints.

A valued constraint language $\Gamma$ is called **tractable** if, for every finite subset $\Gamma' \subseteq \Gamma$, there exists an algorithm solving any instance $\mathcal{P} \in \text{VCSP}(\Gamma')$ in polynomial time. Conversely, $\Gamma$ is called **NP-hard** if there is some finite subset $\Gamma' \subseteq \Gamma$ for which VCSP($\Gamma'$) is NP-hard. Example 2.2 shows that the valued constraint language $\{\varrho_=\}$ is tractable, and Example 2.3 shows that the valued constraint language $\{\varrho_{\neq}\}$ is NP-hard.

We now define a closure operator on weighted relations, which adds to a given set of weighted relations all other weighted relations which can be *expressed* using that set, in the sense defined below.

**Definition 2.5.** *For any VCSP instance $\mathcal{P} = \langle V, D, C \rangle$, and any list $L = \langle v_1, \ldots, v_r \rangle$ of variables of $\mathcal{P}$, the projection of $\mathcal{P}$ onto $L$, denoted $\pi_L(\mathcal{P})$, is the weighted relation on $D$ of arity $r$ defined as follows:*

$$\pi_L(\mathcal{P})(x_1, \ldots, x_r) \stackrel{\text{def}}{=} \min_{\{s:V \to D \mid \langle s(v_1), \ldots, s(v_r) \rangle = \langle x_1, \ldots, x_r \rangle\}} \text{Cost}_{\mathcal{P}}(s).$$

*We say that a weighted relation $\varrho$ is **expressible** over a valued constraint language $\Gamma$ if there exists a VCSP instance $\mathcal{P} \in \text{VCSP}(\Gamma)$ and a list $L$ of variables of $\mathcal{P}$ such that $\pi_L(\mathcal{P}) = \varrho$. We call the pair $\langle \mathcal{P}, L \rangle$ a **gadget** for expressing $\varrho$ over $\Gamma$.*

*We define* Express($\Gamma$) *to be the **expressive power** of $\Gamma$; that is, the set of all weighted relations expressible over $\Gamma$.*

Note that the list of variables $L$ in a gadget may contain repeated entries, the sum over an empty set is zero, and the minimum over an empty set is undefined.

**Example 2.6.** *Let $\mathcal{P}$ be the VCSP instance with a single variable $v$ and no constraints, and let $L = \langle v, v \rangle$. Then, by Definition 2.5,*
$$\pi_L(\mathcal{P})(x, y) = \begin{cases} 0 & \text{if } x = y \\ undefined & \text{otherwise} \end{cases} .$$

*Hence for any valued constraint language $\Gamma$, over any set $D$, $\mathrm{Express}(\Gamma)$ contains this binary weighted relation, which will be called the **weighted equality relation**.*

**Example 2.7.** *Let $\mathcal{P}$ be the VCSP instance with domain $\{0, 1\}$, variables $v_1, v_2, v_3$, and constraints $\langle \langle v_1, v_2 \rangle, \varrho_{\neq} \rangle$ and $\langle \langle v_2, v_3 \rangle, \varrho_{\neq} \rangle$, and let $L = \langle v_1, v_3 \rangle$. Then, by Definition 2.5,*
$$\pi_L(\mathcal{P})(x, y) = \begin{cases} 0 & \text{if } x = y \\ 1 & \text{otherwise} \end{cases} .$$

*Hence $\pi_L(\mathcal{P}) = \varrho_{=}$, so we have that $\varrho_{=} \in \mathrm{Express}(\{\varrho_{\neq}\})$.*

*However, using the results of this paper, we will be able to show, for example, that $\varrho_{\neq} \notin \mathrm{Express}(\{\varrho_{=}\})$ (see Example 8.4).*

The next result shows that expressibility preserves tractability.

**Theorem 2.8.** *A valued constraint language $\Gamma$ is tractable if and only if $\mathrm{Express}(\Gamma)$ is tractable; similarly, $\Gamma$ is NP-hard if and only if $\mathrm{Express}(\Gamma)$ is NP-hard.*

*Proof.* By the definition of a tractable valued constraint language, it is sufficient to show that for any finite subset $\Gamma'$ of $\mathrm{Express}(\Gamma)$ there exists a polynomial-time reduction from $\mathrm{VCSP}(\Gamma')$ to $\mathrm{VCSP}(\Gamma'')$, where $\Gamma''$ is a finite subset of $\Gamma$.

Let $\Gamma'$ be a finite subset of $\mathrm{Express}(\Gamma)$ and let $\mathcal{P}'$ be any instance of $\mathrm{VCSP}(\Gamma')$. By Definition 2.5, any weighted relation $\varrho' \in \mathrm{Express}(\Gamma)$ can be constructed by using some gadget $\langle \mathcal{P}_{\varrho'}, L \rangle$ where $\mathcal{P}_{\varrho'}$ is an instance of $\mathrm{VCSP}(\Gamma)$. Hence we can simply replace each constraint in $\mathcal{P}'$ which has a weighted relation $\varrho'$ not already in $\Gamma$ with the corresponding gadget to obtain an instance $\mathcal{P}$ of $\mathrm{VCSP}(\Gamma)$ which is equivalent to $\mathcal{P}'$. The maximum size of any of the gadgets used is a constant determined by the finite set $\Gamma'$, so this construction can be carried out in polynomial time in the size of $\mathcal{P}'$. □

This result shows that, when trying to identify tractable valued constraint languages, it is sufficient to consider only languages of the form $\mathrm{Express}(\Gamma)$. In the following sections, we will show that such languages can be characterised using certain algebraic properties.

## 3 Classical Constraint Satisfaction

In this section we consider the special case when the weights are all zero.

**Definition 3.1.** *We denote by $\mathbf{R}_D$ the set of all zero-valued weighted relations on a set $D$.*

There is a one-to-one correspondence between the set of zero-valued weighted relations $\mathbf{R}_D$ and the set of all relations over $D$. In this correspondence each weighted relation $\varrho$ in $\mathbf{R}_D$ is associated with the relation $R(\varrho)$ containing precisely those tuples on which $\varrho$ is defined. Similarly, each zero-valued weighted relation in $\mathbf{R}_D$ is associated with the *predicate* which is true for precisely those tuples where the weighted relation is defined. Subsets of $\mathbf{R}_D$ are sometimes referred to as *crisp* constraint languages [16] and $\mathrm{VCSP}(\mathbf{R}_D)$ is equivalent to the classical constraint satisfaction problem, or CSP, where each assignment is either allowed (cost 0) or disallowed (infeasible, or cost undefined).

**Definition 3.2.** *A weighted relation $\varrho$ of arity $r$ can be obtained by **addition** from the weighted relation $\varrho_1$ of arity $s$ and the weighted relation $\varrho_2$ of arity $t$ if $\varrho$ satisfies the identity $\varrho(x_1, \ldots, x_r) = \varrho_1(y_1, \ldots, y_s) + \varrho_2(z_1, \ldots, z_t)$, for some (fixed) choice of $y_1, \ldots, y_s$ and $z_1, \ldots, z_t$ from amongst the $x_1, \ldots, x_r$.*

For zero-valued weighted relations this notion of addition corresponds to performing a relational join operation on the associated relations $R(\varrho_1)$ and $R(\varrho_2)$ [26]. It also corresponds to taking a conjunction of the associated predicates [14]. Moreover, minimising a weighted relation $\varrho \in \mathbf{R}_D$ over one of its arguments corresponds to taking a relational projection of $R(\varrho)$ onto its remaining co-ordinates. It also corresponds to existential quantification of the associated predicate over that argument.

**Definition 3.3.** *A set $\Gamma \subseteq \mathbf{R}_D$ is called a **relational clone** if it contains the weighted equality relation and is closed under addition and minimisation over arbitrary arguments.*

*For each $\Gamma \subseteq \mathbf{R}_D$ we define $\mathrm{RelClone}(\Gamma)$ to be the smallest relational clone containing $\Gamma$.*

It is a straightforward consequence of Definitions 2.5 and 3.3 that the expressive power of a crisp constraint language is given by the smallest relational clone containing it, as the next result indicates.

**Proposition 3.4.** *For any $\Gamma \subseteq \mathbf{R}_D$, $\mathrm{Express}(\Gamma) = \mathrm{RelClone}(\Gamma)$.*

This alternative characterisation for the expressive power of a crisp constraint language was first observed in [27], and used to study the complexity of such languages using tools from universal algebra. We now give a brief summary of this algebraic approach.

For any set $D$, a function $f : D^k \to D$ is called a $k$-ary *operation* on $D$.

**Definition 3.5.** *We denote by $\mathbf{O}_D$ the set of all finitary operations on $D$ and by $\mathbf{O}_D^{(k)}$ the $k$-ary operations in $\mathbf{O}_D$.*

**Definition 3.6.** *The $k$-ary **projections** on $D$ are the operations $e_i^{(k)} : D^k \to D$, $(a_1, \ldots, a_k) \mapsto a_i$.*

**Definition 3.7.** *Let $f \in \mathbf{O}_D^{(k)}$ and $g_1, \ldots, g_k \in \mathbf{O}_D^{(\ell)}$. The **superposition** of $f$ and $g_1, \ldots, g_k$ is the $\ell$-ary operation $f[g_1, \ldots, g_k] : D^\ell \to D$, $(x_1, \ldots, x_\ell) \mapsto f(g_1(x_1, \ldots, x_\ell), \ldots, g_k(x_1 \ldots, x_\ell))$.*

**Definition 3.8.** *A set $F \subseteq \mathbf{O}_D$ is called a **clone** of operations if it contains all the projections on $D$ and is closed under superposition. For each $F \subseteq \mathbf{O}_D$ we define $\mathrm{Clone}(F)$ to be the smallest clone containing $F$.*

We can extend $k$-ary operations to operate on tuples in a natural way, as follows. Let $\mathbf{x}_1, \ldots, \mathbf{x}_k$ be tuples of length $r$ over a set $D$, where each $\mathbf{x}_i = \langle x_{i,1}, x_{i,2}, \ldots, x_{i,r}\rangle$. We can obtain another element of $D^r$ by applying $f$ to the tuples $x_i$ co-ordinatewise, as follows:

$$f(\mathbf{x}_1, \ldots, \mathbf{x}_k) \stackrel{\mathrm{def}}{=} \langle f(x_{1,1}, \ldots, x_{k,1}), f(x_{1,2}, \ldots, x_{k,2}), \ldots, f(x_{1,r}, \ldots, x_{k,r})\rangle.$$

**Definition 3.9.** *Let $\varrho$ be a weighted relation of arity $r$ on a set $D$ and let $f \in \mathbf{O}_D^{(k)}$. We say that $f$ is a **polymorphism** of $\varrho$ if, for any $\mathbf{x}_1, \mathbf{x}_2, \ldots, \mathbf{x}_k \in D^r$ such that $\varrho(\mathbf{x}_i)$ is defined for $i = 1, \ldots, k$, we have that $\varrho(f(\mathbf{x}_1, \mathbf{x}_2, \ldots, \mathbf{x}_k))$ is also defined.*

*If $f$ is a polymorphism of $\varrho$ we say $\varrho$ is **invariant** under $f$.*

**Definition 3.10.** *For any valued constraint language $\Gamma$ over a set $D$, we denote by $\mathrm{Pol}(\Gamma)$ the set of all operations on $D$ which are polymorphisms of all weighted relations $\varrho \in \Gamma$ and by $\mathrm{Pol}^{(k)}(\Gamma)$ the $k$-ary operations in $\mathrm{Pol}(\Gamma)$.*

**Definition 3.11.** *For any $F \subseteq \mathbf{O}_D$, we denote by $\mathrm{Inv}(F)$ the set of all weighted relations in $\mathbf{R}_D$ that are invariant under all operations $f \in F$.*

For any set $D$, the mappings $\mathrm{Pol}$ and $\mathrm{Inv}$ form a Galois connection between $\mathbf{O}_D$ and $\mathbf{R}_D$ [9]. A characterisation of this Galois connection for finite sets $D$ is given by the following two theorems, originally obtained for sets of relations [24, 8].

**Theorem 3.12.** *For any finite set $D$, and any finite $\Gamma \subseteq \mathbf{R}_D$, $\mathrm{Inv}(\mathrm{Pol}(\Gamma)) = \mathrm{RelClone}(\Gamma)$.*

**Theorem 3.13.** *For any finite set $D$, and any finite $F \subseteq \mathbf{O}_D$, $\mathrm{Pol}(\mathrm{Inv}(F)) = \mathrm{Clone}(F)$.*

As with any Galois connection [9], this means that there is a one-to-one correspondence between clones and relational clones. Together with Proposition 3.4, this result shows that the expressive power of any crisp constraint language $\Gamma$ on a finite set $D$ corresponds to a particular clone of operations on $D$. Hence, by Theorem 2.8, the search for tractable crisp constraint languages corresponds to a search for suitable clones of operations [27, 14]. This key observation paved the way for applying deep results from universal algebra in the search for tractable constraint languages [11, 15, 12, 13, 4, 2, 3, 5].

## 4 Weighted Relational Clones

In this section we return to the general case of weighted relations taking arbitrary values in $\mathbb{Q}$ in order to define the notion of a *weighted* relational clone.

**Definition 4.1.** *We denote by $\mathbf{\Phi}_D$ the set of all weighted relations on $D$ taking values in $\mathbb{Q}$ and by $\mathbf{\Phi}_D^{(r)}$ the weighted relations in $\mathbf{\Phi}_D$ of arity $r$.*

We now define a closure operator on weighted relations, which adds to a set of weighted relations all other weighted relations which can be obtained from the original set by non-negative scaling and addition of a constant.

**Definition 4.2.** *A weighted relation $\varrho' \in \mathbf{\Phi}_D$ can be obtained from a weighted relation $\varrho \in \mathbf{\Phi}_D$ by **non-negative scaling** and **addition of constants** if there exist $\alpha, \beta \in \mathbb{Q}$ with $\alpha \geq 0$ such that $\varrho' \equiv \alpha \varrho + \beta$. We denote by $\Gamma_\sim$ the smallest set of weighted relations containing $\Gamma$ which is closed under non-negative scaling and addition of constants.*

The next result shows that adding weighted relations that can be obtained by non-negative scaling and addition of constants preserves tractability.

**Theorem 4.3.** *A valued constraint language $\Gamma$ is tractable if and only if $\Gamma_\sim$ is tractable; similarly, $\Gamma$ is NP-hard if and only if $\Gamma_\sim$ is NP-hard.*

*Proof.* By the definition of tractable valued constraint language, it is sufficient to show that for any finite subset $\Gamma'$ of $\Gamma_\sim$ there exists a polynomial-time reduction from $\mathrm{VCSP}(\Gamma')$ to $\mathrm{VCSP}(\Gamma'')$, where $\Gamma''$ is a finite subset of $\Gamma$.

Let $\Gamma'$ be a finite subset of $\Gamma_\sim$ and let $\mathcal{P}'$ be any instance of $\mathrm{VCSP}(\Gamma')$. By Definition 4.2, any weighted relation $\varrho' \in \Gamma_\sim$ can be obtained by non-negative scaling and addition of constants from some weighted relation $\varrho \in \Gamma$. Hence we can replace each of the constraints $\langle \sigma, \varrho' \rangle$ in $\mathcal{P}'$ with a new constraint $\langle \sigma, \varrho \rangle$, where $\varrho \in \Gamma$ and $\varrho' = \alpha \varrho + \beta$ for some non-negative rational value $\alpha$ and some arbitrary rational constant $\beta$, to obtain an instance $\mathcal{P}$ of $\mathrm{VCSP}(\Gamma)$.

The constant $\beta$ is added to the cost of all assignments and so does not affect the choice of solution.

Since $\alpha$ is a non-negative rational value, it can be expressed as $p/q$ for some non-negative integer $p$ and positive integer $q$.

If $p$ is non-zero, then the effect of the scale factor $p/q$ can be simulated by taking $p$ copies of the new constraint in $\mathcal{P}$ and $q$ copies of all other constraints in $\mathcal{P}$. The maximum values of $p, q$ are constants determined by the finite set $\Gamma'$, so this construction can be carried out in polynomial time in the size of $\mathcal{P}'$.

It only remains to deal with the cases where $p$ is zero. Assume that $\mathcal{P}'$ contains $k$ constraints of the form $\langle \sigma, \varrho' \rangle$ where $\varrho' = 0\varrho + \beta$, and replace each of these with the corresponding constraint $\langle \sigma, \varrho \rangle$ to obtain a new instance, $\mathcal{P}$. Let $M$ be the maximum weight assigned by any weighted relation in the finite set $\Gamma'$, and let $m$ be the minimum difference between any two distinct weights assigned by weighted relations in $\Gamma'$. The cost of any feasible assignment for $\mathcal{P}$ is greater by at most $kM$ than the cost of the same assignment for $\mathcal{P}'$. Hence if we also take $\lceil \frac{Mk}{m} + 1 \rceil$ copies of all the remaining constraints of $\mathcal{P}'$, then we obtain an instance of VCSP($\Gamma$) with the same solutions as $\mathcal{P}'$. Since $M$ and $m$ are constants determined by the finite set $\Gamma'$, this construction can again be carried out in polynomial time in the size of $\mathcal{P}'$. □

**Definition 4.4.** *A set $\Gamma \subseteq \mathbf{\Phi}_D$ is a* **weighted relational clone** *if it contains the weighted equality relation and is closed under non-negative scaling and addition of constants, addition, and minimisation over arbitrary arguments.*

*For each $\Gamma \subseteq \mathbf{\Phi}_D$ we define* wRelClone($\Gamma$) *to be the smallest weighted relational clone containing $\Gamma$.*

It is a straightforward consequence of Definitions 2.5, 4.2 and 4.4 that, for any valued constraint language $\Gamma \subseteq \mathbf{\Phi}_D$, the set of weighted relations that can be expressed using weighted relations obtained from $\Gamma$ by non-negative scaling and addition of constants, is given by the smallest weighted relational clone containing $\Gamma$, as the next result indicates.

**Proposition 4.5.** *For any $\Gamma \subseteq \mathbf{\Phi}_D$,* Express($\Gamma_\sim$) = wRelClone($\Gamma$).

Hence, by Theorem 2.8 and Theorem 4.3, the search for tractable valued constraint languages corresponds to a search for suitable weighted relational clones.

In the next section we establish an alternative characterisation for weighted relational clones which facilitates this search.

## 5 Weighted Clones

To obtain a suitable alternative characterisation for weighted relational clones we now generalise the notion of a clone of operations, introduced in Definition 3.8, by introducing the notion of a *weighted* clone.

Recall that a clone of operations, $C$, is a set of operations on some fixed set $D$ that contains all projections and is closed under superposition. The $k$-ary operations in a clone $C$ will be denoted $C^{(k)}$.

**Definition 5.1.** *We define a $k$-ary* **weighting** *of a clone $C$ to be a function $\omega : C^{(k)} \to \mathbb{Q}$ such that $\omega(f) < 0$ only if $f$ is a projection and*

$$\sum_{f \in C^{(k)}} \omega(f) = 0.$$

*We denote by $\mathbf{W}_C$ the set of all possible weightings of $C$ and by $\mathbf{W}_C^{(k)}$ the set of $k$-ary weightings of $C$.*

For any weighting $\omega$, we denote by $\mathbf{dom}(\omega)$ the set of operations on which $\omega$ is defined. We denote by $\mathrm{ar}(\omega)$ the arity of $\omega$.

Since a weighting is simply a rational-valued function satisfying certain inequalities it can be scaled by any non-negative rational to obtain a new weighting. Similarly, any two weightings of the same clone of the same arity can be added to obtain a new weighting of that clone.

The notion of superposition from Definition 3.7 can also be extended to weightings in a natural way, by forming a superposition with each argument of the weighting, as follows.

**Definition 5.2.** *For any clone $C$, any $\omega \in \mathbf{W}_C^{(k)}$ and any $g_1, g_2, \ldots, g_k \in C^{(\ell)}$, we define the* **superposition** *of $\omega$ and $g_1, \ldots, g_k$, to be the function $\omega[g_1, \ldots, g_k] : C^{(\ell)} \to \mathbb{Q}$ defined by*

$$\omega[g_1, \ldots, g_k](f') \stackrel{\text{def}}{=} \sum_{\substack{f \in C^{(k)} \\ f[g_1, \ldots, g_k] = f'}} \omega(f). \tag{1}$$

**Example 5.3.** *Let $D$ be a totally ordered set $D$, and let $C = \text{Clone}(\{\max\})$ where $\max$ is the binary maximum operation on $D$. Note that $C^{(2)}$ contains just three binary operations: $\text{e}_1^{(2)}, \text{e}_2^{(2)}$ and $\max$.*

*Let $\omega$ be the 2-ary weighting of $C$ given by*

$$\omega(f) \stackrel{\text{def}}{=} \begin{cases} -1 & \text{if } f = \text{e}_1^{(2)} \\ +1 & \text{if } f = \text{e}_2^{(2)} \\ 0 & \text{if } f = \max \end{cases}$$

*and let*

$$\langle g_1, g_2 \rangle = \left\langle \text{e}_2^{(2)}, \max \right\rangle .$$

*Note that $\text{e}_1^{(2)}[g_1, g_2] = g_1 = \text{e}_2^{(2)}$ and $\text{e}_2^{(2)}[g_1, g_2] = g_2 = \max$, so, applying Definition 5.2, we have*

$$\omega[g_1, g_2](f) = \begin{cases} 0 & \text{if } f = \text{e}_1^{(2)} \\ -1 & \text{if } f = \text{e}_2^{(2)} \\ +1 & \text{if } f = \max \end{cases} .$$

*Note that $\omega[g_1, g_2]$ satisfies the conditions of Definition 5.1 and hence is a 2-ary weighting of $C$.*

**Example 5.4.** *Let $C$ be a clone on some totally ordered set $D$ and let $\max$ and $\min$ be binary maximum and minimum operations which are contained in $C$. Note that $C^{(4)}$ contains operations such as $\max[\text{e}_i^{(4)}, \text{e}_j^{(4)}]$ which returns the maximum of the $i$th and $j$th argument values. Operations of this form will be denoted $\max(x_i, x_j)$.*

*Let $\omega$ be the 4-ary weighting of $C$ given by*

$$\omega(f) \stackrel{\text{def}}{=} \begin{cases} -1 & \text{if } f \text{ is a projection, that is, } f \in \{\text{e}_1^{(4)}, \text{e}_2^{(4)}, \text{e}_3^{(4)}, \text{e}_4^{(4)}\} \\ +1 & \text{if } f \in \{\max(x_1, x_2), \min(x_1, x_2), \max(x_3, x_4), \min(x_3, x_4)\} \\ 0 & \text{otherwise} \end{cases}$$

*and let*

$$\langle g_1, g_2, g_3, g_4 \rangle = \left\langle \text{e}_1^{(3)}, \text{e}_2^{(3)}, \text{e}_3^{(3)}, \max(x_1, x_2) \right\rangle .$$

*Then, by Definition 5.2 we have*

$$\omega[g_1, g_2, g_3, g_4](f) = \begin{cases} -1 & \text{if } f \text{ is a projection, that is, } f \in \{\text{e}_1^{(3)}, \text{e}_2^{(3)}, \text{e}_3^{(3)}\} \\ +1 & \text{if } f \in \{\max(x_1, x_2, x_3), \min(x_1, x_2), \min(x_3, \max(x_1, x_2))\} \\ 0 & \text{otherwise} \end{cases} .$$

*Note that $\omega[g_1, g_2, g_3, g_4]$ satisfies the conditions of Definition 5.1 and hence is a 3-ary weighting of $C$.*

**Example 5.5.** *Let $C$ and $\omega$ be the same as in Example 5.4 but now consider*

$$\langle g_1', g_2', g_3', g_4' \rangle = \left\langle \text{e}_1^{(4)}, \max(x_2, x_3), \min(x_2, x_3), \text{e}_4^{(4)} \right\rangle .$$

*By Definition 5.2 we have*

$$\omega[g'_1, g'_2, g'_3, g'_4](f) = \begin{cases} -1 & \text{if } f \in \{e_1^{(4)}, \max(x_2, x_3), \min(x_2, x_3), e_4^{(4)}\} \\ +1 & \text{if } f \in \left\{ \begin{array}{l} \max(x_1, x_2, x_3), \min(x_1, \max(x_2, x_3)), \\ \max(\min(x_2, x_3), x_4), \min(x_2, x_3, x_4) \end{array} \right\} \\ 0 & \text{otherwise} \end{cases}.$$

*Note that $\omega[g'_1, g'_2, g'_3, g'_4]$ does not satisfy the conditions of Definition 5.1 because, for example, we have that $\omega[g'_1, g'_2, g'_3, g'_4](f) < 0$ when $f = \max(x_2, x_3)$, which is not a projection. Hence $\omega[g'_1, g'_2, g'_3, g'_4]$ is not a valid weighting of $C$.*

It follows immediately from Definition 3.7 that the sum of the weights in any superposition $\omega[g_1, \ldots, g_k]$ is equal to the sum of the weights in $\omega$, which is zero, by Definition 5.1. However, as we have seen in Example 5.5, it is not always the case that an arbitrary superposition satisfies the other condition in Definition 5.1, that negative weights are only assigned to projections. Hence we make the following definition:

**Definition 5.6.** *If the result of a superposition is a valid weighting, then that superposition will be called a **proper** superposition.*

**Remark 5.7.** *The superposition of a projection operation and other projection operations is always a projection operation. So, by Definition 5.2, for any clone $C$ and any $\omega \in \mathbf{W}_C^{(k)}$, if $g_1, \ldots, g_k \in C^{(\ell)}$ are projections, then the function $\omega[g_1, \ldots, g_k]$ can take negative values only on projections, and hence is a valid weighting. This means that a superposition with any list of projections is always a proper superposition.*

We are now ready to define *weighted clones*.

**Definition 5.8.** *A **weighted clone**, $W$, is a non-empty set of weightings of some fixed clone $C$ which is closed under non-negative scaling, addition of weightings of equal arity, and proper superposition with operations from $C$. The clone $C$ is called the **support** of $W$.*

**Example 5.9.** *For any clone, $C$, the set $\mathbf{W}_C$ containing all possible weightings of $C$ is a weighted clone with support $C$.*

**Example 5.10.** *For any clone, $C$, the set $\mathbf{W}_C^0$ containing all zero-valued weightings of $C$ is a weighted clone with support $C$. Note that $\mathbf{W}_C^0$ contains exactly one weighting of each possibly arity, which assigns the value 0 to all operations in $C$ of that arity.*

We now establish a link between weightings and weighted relations, which will allow us to link weighted clones and weighted relational clones.

**Definition 5.11.** *Let $\varrho$ be a weighted relation of arity $r$ on some set $D$ and let $\omega$ be a $k$-ary weighting of some clone of operations $C$ on the set $D$.*

*We say that $\omega$ is a **weighted polymorphism** of $\varrho$ if, for any $\mathbf{x}_1, \mathbf{x}_2, \ldots, \mathbf{x}_k \in D^r$ such that $\varrho(\mathbf{x}_i)$ is defined for $i = 1, \ldots, k$, we have that $\varrho(f(\mathbf{x}_1, \mathbf{x}_2, \ldots, \mathbf{x}_k))$ is defined for all $f \in C^{(k)}$, and*

$$\sum_{f \in C^{(k)}} \omega(f) \varrho(f(\mathbf{x}_1, \mathbf{x}_2, \ldots, \mathbf{x}_k)) \leq 0. \tag{2}$$

*If $\omega$ is a weighted polymorphism of $\varrho$ we say $\varrho$ is **improved** by $\omega$.*

Note that, by Definition 3.9, if $\varrho$ is improved by the weighting $\omega \in \mathbf{W}_C^{(k)}$, then every element of $C^{(k)}$ must be a polymorphism of $\varrho$.

**Example 5.12.** *Consider the class of submodular functions [32]. These are precisely the functions $\varrho$ satisfying*

$$\varrho(\min(\mathbf{x}_1, \mathbf{x}_2)) + \varrho(\max(\mathbf{x}_1, \mathbf{x}_2)) - \varrho(\mathbf{x}_1) - \varrho(\mathbf{x}_2) \leq 0.$$

*In other words, the set of submodular functions are the set of weighted relations with a 2-ary weighted polymorphism $\omega_{sub}$, defined by:*

$$\omega_{sub}(f) \stackrel{\text{def}}{=} \begin{cases} -1 & \text{if } f \in \{e_1^{(2)}, e_2^{(2)}\} \\ +1 & \text{if } f \in \{\min(x_1, x_2), \max(x_1, x_2)\} \\ 0 & \text{otherwise} \end{cases}.$$

*Submodular function minimisation is known to be tractable [32].*

**Definition 5.13.** *For any $\Gamma \subseteq \mathbf{\Phi}_D$, we denote by $\mathrm{wPol}(\Gamma)$ the set of all weightings of $\mathrm{Pol}(\Gamma)$ which are weighted polymorphisms of all weighted relations $\varrho \in \Gamma$. The set of k-ary weightings in $\mathrm{wPol}(\Gamma)$ will be denoted $\mathrm{wPol}^{(k)}(\Gamma)$.*

To define a mapping in the other direction, we need to consider the union of the sets $\mathbf{W}_C$ over all clones $C$ on some fixed set $D$, which will be denoted $\mathbf{W}_D$. If we have a set $W \subseteq \mathbf{W}_D$ which may contain weightings of *different* clones over $D$, then we can extend each of these weightings with zeros, as necessary, so that they are weightings of the same clone $C$, given by

$$C = \mathrm{Clone}(\bigcup_{\omega \in W} \mathbf{dom}(\omega)).$$

This set of extended weightings obtained from $W$ will be denoted $\overline{W}$. For any set $W \subseteq \mathbf{W}_D$, we define $\mathrm{wClone}(W)$ to be the smallest weighted clone containing $\overline{W}$.

**Definition 5.14.** *For any $W \subseteq \mathbf{W}_D$, we denote by $\mathrm{Imp}(W)$ the set of all weighted relations in $\mathbf{\Phi}_D$ which are improved by all weightings $\omega \in W$. The set of r-ary weighted relations in $\mathrm{Imp}(W)$ will be denoted $\mathrm{Imp}^{(r)}(W)$.*

It follows immediately from the definition of a Galois connection [9] that, for any set $D$, the mappings $\mathrm{wPol}$ and $\mathrm{Imp}$ form a Galois connection between $\mathbf{W}_D$ and $\mathbf{\Phi}_D$, as illustrated in Figure 1. A characterisation of this Galois connection for finite sets $D$ is given by the following two theorems, which are proved in Section 6.

**Theorem 5.15.** *For any finite set $D$, and any finite $\Gamma \subseteq \mathbf{\Phi}_D$, $\mathrm{Imp}(\mathrm{wPol}(\Gamma)) = \mathrm{wRelClone}(\Gamma)$.*

**Theorem 5.16.** *For any finite set $D$, and any finite $W \subseteq \mathbf{W}_D$, $\mathrm{wPol}(\mathrm{Imp}(W)) = \mathrm{wClone}(W)$.*

As with any Galois connection [9], this means that there is a one-to-one correspondence between weighted clones and weighted relational clones. Hence, by Proposition 4.5, Theorem 2.8, and Theorem 4.3, the search for tractable valued constraint languages over a finite set corresponds to a search for suitable weighted clones of operations.

# 6 Proofs of Theorems 5.15 and 5.16

Our proofs of Theorems 5.15 and 5.16 will both use the following result, which is a variant of the well-known Farkas' Lemma used in linear programming [32, 35].

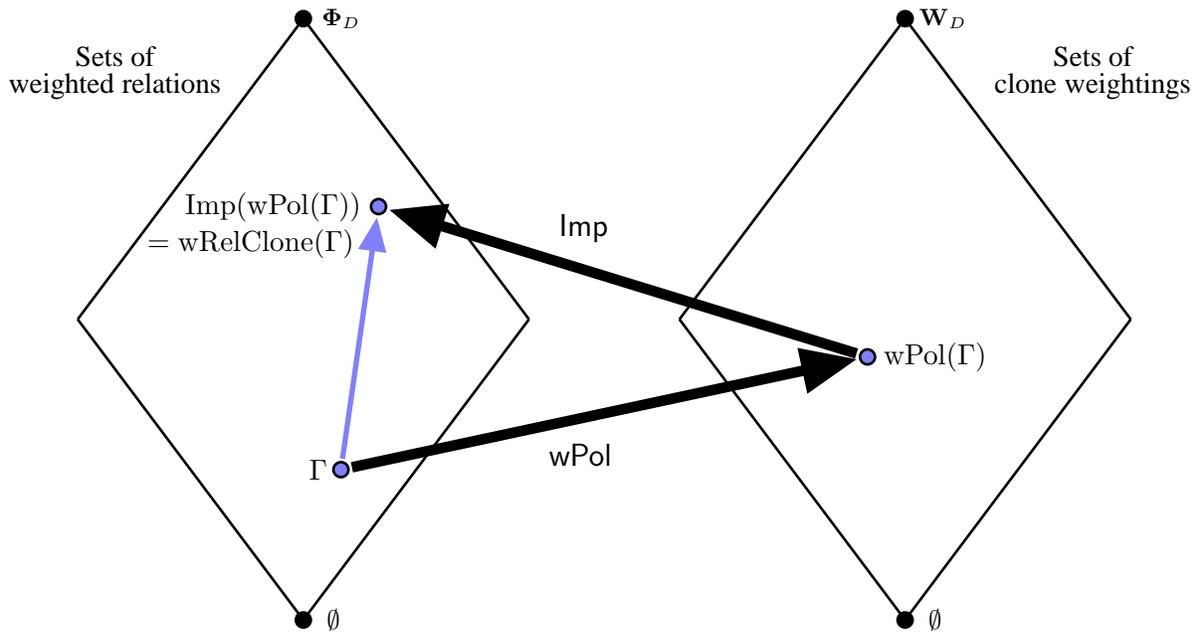

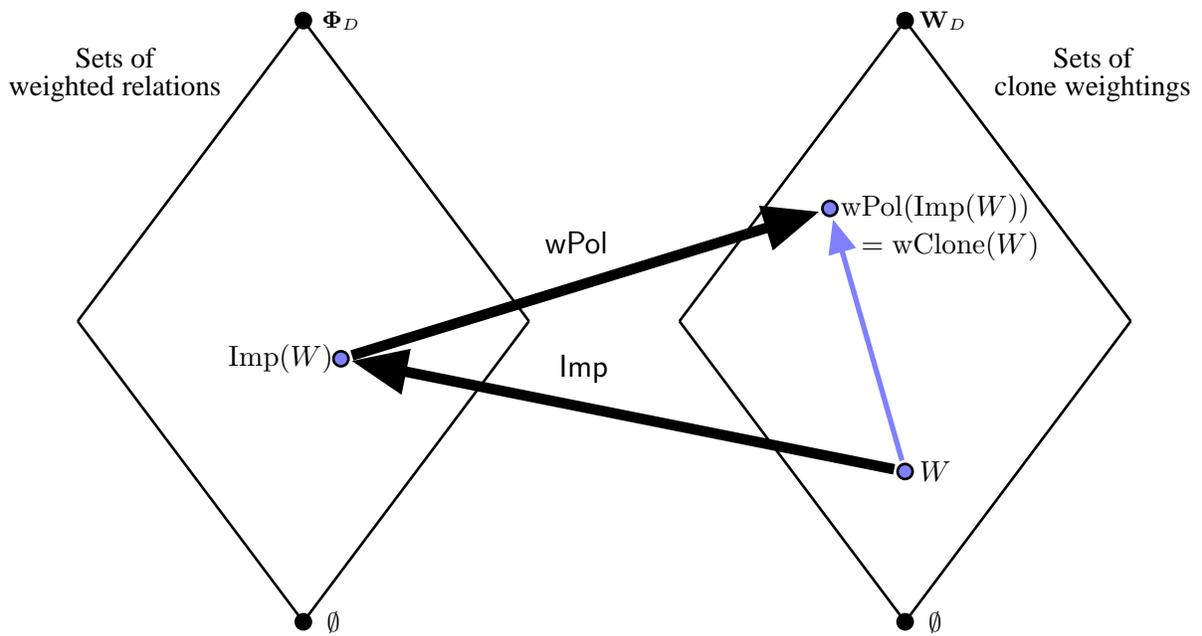

Figure 1: Galois connection between $\mathbf{\Phi}_D$ and $\mathbf{W}_D$.

**Lemma 6.1** (Farkas 1894). *Let $S$ and $T$ be finite sets of indices, where $T$ is the disjoint union of two subsets, $T_\geq$ and $T_=$. For all $i \in S$, and all $j \in T$, let $a_{i,j}$ and $b_j$ be rational numbers. Exactly one of the following holds:*

- *Either there exists a set of non-negative rational numbers $\{x_i \mid i \in S\}$ and a rational number $C$ such that*

$$\text{for each } j \in T_\geq, \quad \sum_{i \in S} a_{i,j} \, x_i \;\geq\; b_j + C, \quad \text{and,}$$

$$\text{for each } j \in T_=, \quad \sum_{i \in S} a_{i,j} \, x_i \;=\; b_j + C.$$

- *Or else there exists a set of integers $\{y_j \mid j \in T\}$ such that $\sum_{j \in T} y_j = 0$ and:*

$$\text{for each } j \in T_\geq, \quad y_j \;\geq\; 0,$$

$$\text{for each } i \in S, \quad \sum_{j \in T} y_j \, a_{i,j} \;\leq\; 0, \quad \text{and}$$

$$\sum_{j \in T} y_j \, b_j \;>\; 0.$$

*Such a set is called a **certificate of unsolvability**.*

We note that there is an effective procedure to decide which of the cases mentioned in Lemma 6.1 holds for any instance, and to calculate the values of the corresponding coefficients $x_i$ or $y_j$ [35].

We will prove Theorem 5.15 in two parts. First, we show in Proposition 6.2 that the set of all weighted relations improved by any given set of weightings is always a weighted relational clone. Then we show that for any finite set $\Gamma$ the set of weighted relations improved by all weightings in $\mathrm{wPol}(\Gamma)$ is precisely the weighted relational clone $\mathrm{wRelClone}(\Gamma)$.

**Proposition 6.2.** *For any finite set $D$, and any $W \subset \mathbf{W}_D$, $\mathrm{Imp}(W)$ is a weighted relational clone.*

*Proof.* Certainly $\mathrm{Imp}(W)$ contains the weighted equality relation, $\varrho_=$, since $\varrho_=$ satisfies inequality (2) in Definition 5.11 for all $\mathbf{x}_i$ such that $\varrho_=(\mathbf{x}_i)$ is defined. Similarly, $\mathrm{Imp}(W)$ is closed under non-negative scaling, addition of constants, addition and rearrangement of arguments, since all of these operations preserve inequality (2). Hence, to show $\mathrm{Imp}(W)$ is a weighted relational clone we only need to show $\mathrm{Imp}(W)$ is closed under minimisation.

Let $\varrho \in \mathrm{Imp}^{(r)}(W)$ and assume that $\varrho'$ is obtained from $\varrho$ by minimising over the last argument. In other words, $\varrho'(x_1, x_2, \ldots, x_{r-1}) = \min_{x_r}(\varrho(x_1, x_2, \ldots, x_r))$. We will now show that $\varrho' \in \mathrm{Imp}(W)$.

Let $\omega \in W$ be a $k$-ary weighting of a clone $C$. Since $\varrho \in \mathrm{Imp}(W)$, we know that $\varrho$ and $\omega$ satisfy inequality (2) for all $\mathbf{x}_1, \mathbf{x}_2, \ldots, \mathbf{x}_k$ such that $\varrho(\mathbf{x}_i)$ is defined. Now consider any $\mathbf{x}'_1, \mathbf{x}'_2, \ldots, \mathbf{x}'_k$ for which each $\varrho'(\mathbf{x}'_i)$ is defined. Extend each $\mathbf{x}'_i$ to a tuple $\mathbf{x}''_i$ of arity $r$ in such a way that $\varrho(\mathbf{x}''_i)$ is minimised. Since all negative values of $\omega$ are associated with projections, we have

$$\sum_{f \in C^{(k)}} \omega(f) \varrho'(f(\mathbf{x}'_1, \mathbf{x}'_2, \ldots, \mathbf{x}'_k)) \leq \sum_{f \in C^{(k)}} \omega(f) \varrho(f(\mathbf{x}''_1, \mathbf{x}''_2, \ldots, \mathbf{x}''_k)) \leq 0 \,.$$

□

We now prove Theorem 5.15, which states that that for any finite set $D$, and any finite $\Gamma \subset \mathbf{\Phi}_D$, $\mathrm{Imp}(\mathrm{wPol}(\Gamma)) = \mathrm{wRelClone}(\Gamma)$.

*Proof.* We first establish that for any $\Gamma \subset \mathbf{\Phi}_D$ we have the inclusion $\text{wRelClone}(\Gamma) \subseteq \text{Imp}(\text{wPol}(\Gamma))$. To see this, observe that $\Gamma \subseteq \text{Imp}(\text{wPol}(\Gamma))$ and, hence, $\text{wRelClone}(\Gamma) \subseteq \text{wRelClone}(\text{Imp}(\text{wPol}(\Gamma)))$ which is equal to $\text{Imp}(\text{wPol}(\Gamma))$ by Proposition 6.2.

We will prove the reverse inclusion, $\text{Imp}(\text{wPol}(\Gamma)) \subseteq \text{wRelClone}(\Gamma)$, as follows. Given a weighted relation $\varrho$ of arity $r$, we will show that either there exists a weighted operation $\omega \in \text{wPol}(\Gamma)$ such that $\varrho \notin \text{Imp}(\{\omega\})$ or else $\varrho$ plus some constant can be obtained by minimisation from a non-negative weighted sum of weighted relations in $\Gamma$, and hence $\varrho \in \text{wRelClone}(\Gamma)$.

We now give the details of this argument. Let $k$ be the number of $r$-tuples for which $\varrho$ is defined and fix an arbitrary order, $\mathbf{x}_1, \ldots, \mathbf{x}_k$, for these tuples. This list of tuples can be viewed as (the rows of) a matrix with $k$ rows and $r$ columns, which we will call $S_\varrho$.

By Proposition 4.5, $\varrho \in \text{wRelClone}(\Gamma)$ if and only if it can be expressed using weighted relations from $\Gamma_\sim$. By Definition 2.5, a weighted relation $\varrho'$ is expressible over $\Gamma_\sim$ if and only if there exists an instance $\mathcal{P} \in \text{VCSP}(\Gamma_\sim)$ and a list $L$ of variables of $\mathcal{P}$ such that $\pi_L(\mathcal{P}) = \varrho'$.

We consider instances $\mathcal{P}$ with $|D^k|$ variables, where each variable is associated with a distinct tuple from $D^k$. Each constraint of $\mathcal{P}$ is a pair $\langle S, \gamma \rangle$ for some $\gamma \in \Gamma_\sim$ and some list of variables $S$. Each such $S$ can be viewed as a list of $k$-tuples over $D$, and hence as a matrix over $D$, whose columns are these $k$-tuples. Since we are using $\mathcal{P}$ to express the defined values of $\varrho$, it is sufficient to consider only matrices $S$ with rows $\mathbf{t}_1, \ldots, \mathbf{t}_k$ such that $\gamma(\mathbf{t}_i)$ is defined for $i = 1, \ldots, k$. For any $\gamma \in \Gamma$, a pair $\langle S, \gamma \rangle$ with this property will be called a $k$-match to $\Gamma$.

Each assignment to the variables of $\mathcal{P}$ can be seen as a mapping from $k$-tuples over $D$ to $D$, and hence associated with an operation $f \in \mathbf{O}_D^{(k)}$. For any list of variables $S$ of $\mathcal{P}$, we will write $f(S)$ to denote the assignment to those variables obtained by applying $f$ to the columns of $S$, viewed as a matrix. With this notation, we have that $\pi_L(\mathcal{P}) = \varrho'$ with $L = S_\varrho$ if we can find non-negative rationals $x_{S,\gamma}$ for all $k$-matches to $\Gamma$, and a constant $c \in \mathbb{Q}$, such that the following system of inequalities and equations is satisfied:

For each $f \in \mathbf{O}_D^{(k)}$,

$$\sum_{\gamma \in \Gamma} \sum_{\{\text{all } k\text{-matches } \langle S, \gamma \rangle\}} x_{S,\gamma}\, \gamma(f(S)) \quad \geq \quad \varrho(f(S_\varrho)) + c$$

For each projection $e \in \mathbf{O}_D^{(k)}$,

$$\sum_{\gamma \in \Gamma} \sum_{\{\text{all } k\text{-matches } \langle S, \gamma \rangle\}} x_{S,\gamma}\, \gamma(e(S)) \quad = \quad \varrho(e(S_\varrho)) + c$$

Moreover, if $f \notin \text{Pol}^{(k)}(\Gamma)$ then the left-hand-side of the corresponding inequality will be undefined, by Definition 3.9, so it is sufficient to consider only $f \in \text{Pol}^{(k)}(\Gamma)$. This gives us a system of inequalities and equations with rational coefficients.

If this system has a solution then $\varrho \in \text{wRelClone}(\Gamma)$. On the other hand, if this system of equations and inequalities has no solution, then we appeal to Lemma 6.1, to get a certificate of unsolvability. That is, in this case we know that there exists a set of integers $\{y_f \mid f \in \text{Pol}^{(k)}(\Gamma)\}$, such that $\sum_{f \in \text{Pol}^{(k)}(\Gamma)} y_f = 0$, $y_f < 0$ only if $f$ is a projection, and:

$$\text{for each } k\text{-matching } \langle S, \gamma \rangle \text{ of } \Gamma, \quad \sum_{f \in \text{Pol}^{(k)}(\Gamma)} y_f\, \gamma(f(S)) \leq 0, \quad \text{and} \tag{3}$$

$$\sum_{f \in \text{Pol}^{(k)}(\Gamma)} y_f\, \varrho(f(S_\varrho)) > 0 \tag{4}$$

Now, consider the $k$-ary weighting $\omega$ of the clone $\mathrm{Pol}(\Gamma)$ defined by $\omega(f) = y_f$ for each $f \in \mathrm{Pol}^{(k)}(\Gamma)$. From (3), we can see that $\omega$ is a weighted polymorphism of every $\gamma \in \Gamma$. On the other hand, (4) shows that $\omega$ is not a weighted polymorphism of $\varrho$. □

**Remark 6.3.** *The proof of Theorem 5.15 demonstrates the decidability of the following question: for any finite $\Gamma \subseteq \mathbf{\Phi}_D$ and any weighted relation $\varrho$ defined on $D$, does $\varrho$ belong to* $\mathrm{wRelClone}(\Gamma)$*?*

We will prove Theorem 5.16 in two parts. First, we show in Proposition 6.4 that the set of weighted polymorphisms of any given set of weighted relations is always a weighted clone. Then we show that for any finite set $W$ the set of weightings that improve all weighted relations in $\mathrm{Imp}(W)$ is precisely the weighted clone $\mathrm{wClone}(W)$.

**Proposition 6.4.** *For any finite set $D$, and any $\Gamma \subset \mathbf{\Phi}_D$, $\mathrm{wPol}(\Gamma)$ is a weighted clone.*

*Proof.* By Definition 5.13, $\mathrm{wPol}(\Gamma)$ is a set of weightings of $\mathrm{Pol}(\Gamma)$. Similarly, $\mathrm{wPol}(\Gamma)$ is closed under addition and non-negative scaling, since both of these operations preserve inequality (2). Hence, to show $\mathrm{wPol}(\Gamma)$ is a weighted clone we only need to show $\mathrm{wPol}(\Gamma)$ is closed under proper superposition by members of $\mathrm{Pol}(\Gamma)$.

Let $\omega \in \mathrm{wPol}^{(k)}(\Gamma)$ and suppose $\omega' = \omega[g_1, \ldots, g_k]$ is a proper superposition of $\omega$, where $g_1, g_2, \ldots, g_k \in \mathrm{Pol}^{(\ell)}(\Gamma)$. We will now show that $\omega' \in \mathrm{wPol}^{(\ell)}(\Gamma)$. Suppose $\varrho$ is a weighted relation of arity $r$ satisfying $\omega \in \mathrm{wPol}(\{\varrho\})$, i.e., $\varrho$ and $\omega$ satisfy inequality (2) for all $\mathbf{x}_1, \mathbf{x}_2, \ldots, \mathbf{x}_k$ such that each $\varrho(\mathbf{x}_i)$ is defined. Given any $\mathbf{x}'_1, \mathbf{x}'_2, \ldots, \mathbf{x}'_\ell$ for which each $\varrho(\mathbf{x}'_i)$ is defined, set $\mathbf{x}_i = g_i(\mathbf{x}'_1, \mathbf{x}'_2, \ldots, \mathbf{x}'_\ell)$ for $i = 1, 2, \ldots, k$. Then, if we set $f' = f[g_1, \ldots, g_k]$, we have $f'(\mathbf{x}'_1, \mathbf{x}'_2, \ldots, \mathbf{x}'_\ell) = f(\mathbf{x}_1, \mathbf{x}_2, \ldots, \mathbf{x}_k)$, for any $f \in \mathrm{Pol}^{(k)}(\Gamma)$. Hence, by Definition 5.2, we have

$$\sum_{f' \in \mathrm{Pol}^{(\ell)}(\Gamma)} \omega'(f')\varrho(f'(\mathbf{x}'_1, \mathbf{x}'_2, \ldots, \mathbf{x}'_\ell)) = \sum_{f \in \mathrm{Pol}^{(\ell)}(\Gamma)} \omega(f)\varrho(f(\mathbf{x}_1, \mathbf{x}_2, \ldots, \mathbf{x}_k)) \leq 0.$$

□

We will make use of the following technical lemma, which shows that any weighted sum of arbitrary superpositions of a pair of weightings $\omega_1$ and $\omega_2$ can be obtained by taking a weighted sum of superpositions of $\omega_1$ and $\omega_2$ with projection operations, and then taking a superposition of the result. This result implies that any weighting which can be expressed as a weighted sum of arbitrary superpositions can also be expressed as a superposition of a weighted sum of *proper* superpositions.

**Lemma 6.5.** *Let $C$ be a clone, and let $\omega_1$ and $\omega_2$ be weightings of $C$, of arity $k$ and $\ell$ respectively. For any $g_1, \ldots, g_k \in C^{(m)}$ and any $g'_1, \ldots, g'_\ell \in C^{(m)}$,*

$$c_1 \, \omega_1[g_1, \ldots, g_k] + c_2 \, \omega_2[g'_1, \ldots, g'_\ell] = \omega[g_1, \ldots, g_k, g'_1, \ldots, g'_\ell], \tag{5}$$

*where $\omega = c_1 \, \omega_1[e_1^{(k+\ell)}, \ldots, e_k^{(k+\ell)}] + c_2 \, \omega_2[e_{k+1}^{(k+\ell)}, \ldots, e_{k+\ell}^{(k+\ell)}]$*

*Proof.* For any $f \in C^{(m)}$, the result of applying the right-hand side expression in equation (5) to $f$ is:

$$\sum_{\substack{f' \in C^{(k+\ell)} \\ f'[g_1, \ldots, g_k, g'_1, \ldots, g'_\ell] = f}} \left( \sum_{\substack{h' \in C^{(k)} \\ h'[e_1^{(k+\ell)}, \ldots, e_k^{(k+\ell)}] = f'}} c_1 \, \omega_1(h') + \sum_{\substack{h' \in C^{(\ell)} \\ h'[e_{k+1}^{(k+\ell)}, \ldots, e_{k+\ell}^{(k+\ell)}] = f'}} c_2 \, \omega_2(h') \right).$$

Replacing each $f'$ by the equivalent superposition of $h'$ with projections, we obtain:

$$\sum_{\substack{h'\in C^{(k)} \\ h'[g_1,\ldots,g_k]=f}} c_1\,\omega_1(h') + \sum_{\substack{h'\in C^{(\ell)} \\ h'[g'_1,\ldots,g'_\ell]=f}} c_2\,\omega_2(h')\,,$$

which is the result of applying the left-hand-side of Equation 5 to $f$. □

We now prove Theorem 5.16, which states that for any finite set $D$, and any finite $W \subset \mathbf{W}_D$, $\mathrm{wPol}(\mathrm{Imp}(W)) = \mathrm{wClone}(W)$.

*Proof.* We first establish that for any $W \subset \mathbf{W}_D$ we have the inclusion $\mathrm{wClone}(W) \subseteq \mathrm{wPol}(\mathrm{Imp}(W))$. To see this, observe that every operation in $C = \mathrm{Clone}(\bigcup_{\omega\in W} \mathbf{dom}(\omega))$ is a polymorphism of $\mathrm{Imp}(W)$, by Definition 3.9, so $\overline{W} \subseteq \mathrm{wPol}(\mathrm{Imp}(W))$. Hence, $\mathrm{wClone}(W) \subseteq \mathrm{wClone}(\mathrm{wPol}(\mathrm{Imp}(W)))$ which is equal to $\mathrm{wPol}(\mathrm{Imp}(W))$ by Proposition 6.4.

We will prove the reverse inclusion, $\mathrm{wPol}(\mathrm{Imp}(W)) \subseteq \mathrm{wClone}(W)$, as follows. Given any weighting $\omega_0 \in \mathbf{W}_D$, we will show that either there exists a weighted relation $\varrho \in \mathrm{Imp}(W)$ such that $\omega_0 \notin \mathrm{wPol}(\{\varrho\})$ or else $\omega_0$ is equal to a non-negative weighted sum of superpositions of weightings in $W$, and hence $\omega_0 \in \mathrm{wClone}(W)$.

We now give the details of this argument. Let $k$ be the arity of $\omega_0$, and let $M = |D|^k$. We first observe that it is sufficient to consider weighted relations of arity $M$ in $\mathrm{Imp}(W)$. To see this, suppose there exists a weighted relation $\varrho \in \mathrm{Imp}(W)$ with arity $N > M$ such that $\omega_0 \notin \mathrm{wPol}(\{\varrho\})$ and let $\mathbf{x}_1, \ldots, \mathbf{x}_k \in D^N$ be any set of tuples for which inequality (2) fails to hold for $\omega_0$ and $\varrho$. Let $\mathbf{X}$ be the $k \times N$ matrix whose rows are the tuples $\mathbf{x}_1, \ldots, \mathbf{x}_k$. Since $N > M$ it follows that some of the columns in this matrix must be equal. Moreover, if the $i$-th and $j$-th column of $\mathbf{X}$ are equal, then so will be the $i$-th and $j$-th entry of the tuple $f(\mathbf{x}_1, \ldots, \mathbf{x}_k)$ obtained by applying any $f \in \mathbf{O}_D^{(k)}$ to these $k$ tuples.

Now let $\varrho'$ be the weighted relation of arity $\leq M$ that depends only on the first of each of these repeated columns, and takes the same values as $\varrho$ takes on arguments with the appropriate entries repeated. Let $\mathbf{X}'$ be the reduced form of $\mathbf{X}$ (with repeated columns deleted). By this approach, we can construct $\varrho'$ so that $\varrho' \in \mathrm{Imp}(W)$, but $\mathbf{X}'$ gives a certificate to show that $\omega_0 \notin \mathrm{wPol}(\{\varrho\})$, i.e., the rows of $\mathbf{X}'$ form a list of tuples for which (2) fails to hold for $\omega_0$ and $\varrho'$.

Moreover, if we have a weighted relation $\varrho \in \mathrm{Imp}(W)$ with arity $N < M$ such that $\omega_0 \notin \mathrm{wPol}(\{\varrho\})$, then $\varrho$ can be extended to a weighted relation $\varrho'$ of arity $M$ that does not depend on the $M - N$ added inputs, and, hence, is also contained in $\mathrm{Imp}(W)$ and is such that $\omega_0 \notin \mathrm{wPol}(\{\varrho'\})$.

By the argument given above, there exists a weighted relation $\varrho \in \mathrm{Imp}(W)$ such that $\omega_0 \notin \mathrm{wPol}(\{\varrho\})$ if and only if there exists a weighted relation $\varrho_M$ of arity $M$ in $\mathrm{Imp}(W)$ such that $\omega_0 \notin \mathrm{wPol}(\{\varrho_M\})$. Furthermore, by reordering the arguments of $\varrho_M$ if necessary, we can assume that $\varrho_M$ and $\omega_0$ violate (2) on the particular list of tuples $\mathbf{x}_1, \ldots, \mathbf{x}_k$ given by taking the rows of a matrix, $\mathbf{X}_M$, whose columns are precisely the tuples in $D^k$, ordered lexicographically.

By Definition 5.11, such a weighted relation $\varrho_M$ exists if and only if the following system of inequalities can be satisfied, for all $\omega \in W$ and all $\mathbf{t}_1, \ldots, \mathbf{t}_{\mathrm{ar}(\omega)} \in D^M$ such that $\varrho_M(\mathbf{t}_i)$ is defined for $i = 1, \ldots, \mathrm{ar}(\omega)$,

$$\sum_{g\in\mathbf{dom}(\omega)} \omega(g)\,\varrho_M(g(\mathbf{t}_1,\ldots,\mathbf{t}_{\mathrm{ar}(\omega)})) \;\leq\; 0\,, \tag{6}$$

and, for the tuples $\mathbf{x}_1, \ldots, \mathbf{x}_k$ forming the rows of $\mathbf{X_M}$, $\varrho_M(\mathbf{x}_i)$ is defined for $i = 1, \ldots, k$ and

$$\sum_{f\in\mathbf{dom}(\omega_0)} \omega_0(f)\,\varrho_M(f(\mathbf{x}_1,\ldots,\mathbf{x}_k)) \;\not\leq\; 0\,. \tag{7}$$

There is a one-to-one correspondence between operations $g : D^k \to D$ and tuples $\mathbf{t}_g \in D^M$, where the tuple $\mathbf{t}_g$ contains the list of values returned by the operation $g$ applied to the columns of $\mathbf{X_M}$.

Set $C = \text{Clone}(\bigcup_{\omega \in W} \mathbf{dom}(\omega))$. We observe that, to satisfy inequality (6), for any $\omega \in W$, if $\varrho_M(\mathbf{t}_i)$ is defined for $i = 1, \ldots, \text{ar}(\omega)$, then $\varrho_M(g(\mathbf{t}_1, \ldots, \mathbf{t}_{\text{ar}(\omega)}))$ must be defined for all $g \in \mathbf{dom}(\omega)$. To achieve this, it is sufficient to ensure that, for all $g \in C^{(k)}$, $\varrho_M(\mathbf{t}_g)$ is defined. All other values of $\varrho_M$ can be left undefined, as this just reduces the number of inequalities in the system.

Using superposition (Definition 5.2), we can rewrite inequalities (6) to obtain the following equivalent system: for all $\omega \in \overline{W}$, and all $g_1, \ldots, g_{\text{ar}(\omega)} \in C^{(k)}$,

$$\sum_{f \in \mathbf{dom}(\omega[g_1,\ldots,g_{\text{ar}(\omega)}])} \omega[g_1, \ldots, g_{\text{ar}(\omega)}](f) \, \varrho_M(f(\mathbf{x}_1, \ldots, \mathbf{x}_k)) \leq 0. \quad (8)$$

Now, by applying Lemma 6.1 to the resulting system of inequalities, we conclude that either a solution $\varrho_M$ exists, in which case $\omega_0 \notin \text{wPol}(\text{Imp}(W))$, or else there exists a set of non-negative rational numbers

$$\{c_{\omega[g_1,\ldots,g_{\text{ar}(\omega)}]} \mid \omega \in \overline{W}, g_1, \ldots, g_{\text{ar}(\omega)} \in C^{(k)}\}$$

such that for every $f \in C^{(k)}$,

$$\sum_{\omega \in \overline{W}} \sum_{\substack{\langle g_1,\ldots,g_{\text{ar}(\omega)}\rangle \\ g_i \in C^{(k)}}} c_{\omega[g_1,\ldots,g_{\text{ar}(\omega)}]} \omega[g_1, \ldots, g_{\text{ar}(\omega)}](f) \geq \omega_0(f). \quad (9)$$

By Definition 5.1, adding the left-hand side of these inequalities over all $f$ gives 0, and so does adding the right hand sides, so each inequality must actually be an equality. In other words, $\omega_0$ is equal to a non-negative weighted sum of superpositions of weightings in $\overline{W}$. Hence, by Lemma 6.5 and Remark 5.7, $\omega_0$ is equal to a proper superposition of some element $\omega_0' \in \text{wClone}(W)$, so $\omega_0 \in \text{wClone}(W)$. $\square$

**Remark 6.6.** *The proof of Theorem 5.16 demonstrates the decidability of the following question: for any finite $W \subseteq \mathbf{W}_D$ and any weighting $\omega$ defined on $D$, does $\omega$ belong to* $\text{wClone}(W)$*?*

## 7 Necessary Conditions For Tractability

In this section, we will start to investigate the structure of weighted clones, and hence establish some necessary conditions for any valued constraint language to be tractable.

Note that, by Definition 3.8, the smallest possible clone of operations over a fixed set $D$ is the set of all projection operations on $D$, which is denoted $\mathbf{J}_D$.

**Proposition 7.1.** *For any finite set $D$, there are precisely two weighted clones with support $\mathbf{J}_D$. These are $\mathbf{W}_{\mathbf{J}_D}$ and $\mathbf{W}^0_{\mathbf{J}_D}$.*

*Proof.* Let $W$ be a weighted clone with support $\mathbf{J}_D$.

If the weights assigned by every weighting $\omega \in W$ are all zero, then $W$ is the zero-valued weighted clone $\mathbf{W}^0_{\mathbf{J}_D}$ described in Example 5.10.

Otherwise, there is some $\omega \in W$ (of arity $k$) such that $\omega$ assigns positive weight to some $k$-ary projections and negative weights to some of the others (the sum of the weights is zero, by Definition 5.1). If we form the superposition of $\omega$ with the sequence of projections $g_1, g_2, \ldots, g_k$, where $g_i = e_a^{(k)}$ if $\omega(e_i^{(k)})$ is positive, and $g_i = e_b^{(k)}$ otherwise, then we obtain a new weighting $\omega[g_1, g_2, \ldots, g_k]$ of $\mathbf{J}_D$ which assigns some positive weight $w$ to $e_a^{(k)}$ and $-w$ to $e_b^{(k)}$.

By adding appropriate multiples of such functions for each successive pair of indices $a$ and $b$, we can obtain any desired weighting of $\mathbf{J}_D$. Hence, in this case $W$ contains all possible weightings of $\mathbf{J}_D$, so $W = \mathbf{W}_{\mathbf{J}_D}$. □

Any weighting $\omega$ which is defined only for projection operations will be called a **trivial** weighting,

**Proposition 7.2.** *For any set of trivial weightings $W$, $\mathrm{Imp}(W)$ is NP-hard.*

*Proof.* By Theorems 5.15 and 5.16, we have that $\mathrm{Imp}(\mathrm{wClone}(W)) = \mathrm{wRelClone}(\mathrm{Imp}(W))$, so by by Proposition 4.5, Theorem 2.8, and Theorem 4.3 we have that $\mathrm{Imp}(W)$ is NP-hard if and only if $\mathrm{Imp}(\mathrm{wClone}(W))$ is NP-hard.

If $W$ contains only trivial weightings, then $\mathrm{wClone}(W)$ has support $\mathbf{J}_D$, so it is equal to $\mathbf{W}_{\mathbf{J}_D}$ or $\mathbf{W}^0_{\mathbf{J}_D}$ by Proposition 7.1.

Every weighting in $\mathbf{W}^0_{\mathbf{J}_D}$ is a weighted polymorphism of any possible weighted relation, by Definition 5.11. Hence $\mathrm{Imp}(\mathbf{W}^0_{\mathbf{J}_D}) = \mathbf{\Phi}_D$.

The weighted relations that are improved by *all* weightings are precisely those which take at most one value. Hence $\mathrm{Imp}(\mathbf{W}_{\mathbf{J}_D}) = (\mathbf{R}_D)_\sim$.

In both cases the resulting valued constraint language is NP-hard. □

Now consider weightings whose values are all 0.

**Proposition 7.3.** *For any set of zero-valued weightings $W$, $\mathrm{Imp}(W)$ is NP-hard.*

*Proof.* By Definition 5.11, a zero-valued weighting will be a weighted polymorphism of *any* weighted relation which is a total function (i.e., any weighted relation where all assignments are feasible). Some valued constraint languages containing only total functions are NP-hard [16]. For example, consider the valued constraint language consisting of the following total function:

$$\varrho_{\neq}(x,y) = \begin{cases} 0 & \text{if } x \neq y \\ 1 & \text{otherwise} \end{cases}.$$

We observed in Example 2.3 that on the domain $\{0,1\}$ the problem $\mathrm{VCSP}(\{\varrho_{\neq}\})$ corresponds to the MAX-CUT problem which is known to be NP-hard. Over domains of size $k > 2$ this problem corresponds to the problem MAX-$k$-CUT, which is also known to be NP-hard. □

Using the Galois connection developed in the previous sections, these two results tell us that any valued constraint language that is *not* NP-hard must have a weighted polymorphism which is non-trivial and assigns at least some non-zero weights. A weighting which assigns positive weight to at least one operation that is not a projection will be called a **positive weighting**.

**Corollary 7.4.** *For any finite set $D$, and any $\Gamma \subset \mathbf{\Phi}_D$, either $\Gamma$ is NP-hard, or else $\mathrm{wPol}(\Gamma)$ is a weighted clone containing some positive weightings.*

*Proof.* By Proposition 6.4, in all cases $\mathrm{wPol}(\Gamma)$ is a weighted clone.

By Theorem 5.15, for any finite $\Gamma' \subseteq \Gamma$, $\mathrm{Imp}(\mathrm{wPol}(\Gamma')) = \mathrm{wRelClone}(\Gamma')$. By Proposition 4.5, Theorem 2.8, and Theorem 4.3, if $\Gamma'$ is NP-hard, then $\mathrm{wRelClone}(\Gamma')$ is also NP-hard, so $\mathrm{Imp}(\mathrm{wPol}(\Gamma'))$ must be NP-hard.

Conversely, if $\Gamma'$ is not NP-hard, then the same argument shows that $\mathrm{Imp}(\mathrm{wPol}(\Gamma'))$ is not NP-hard, so by Propositions 7.2 and 7.3, $\mathrm{wPol}(\Gamma')$ must contain some weightings that are non-trivial and some weightings that are not zero-valued.

Choose a weighting $\omega \in \text{wPol}(\Gamma)$ that is not zero-valued, and a weighting $\omega'$ that is non-trivial (but may be zero-valued). If $\omega$ assigns positive weight to any non-projection, then it is a positive weighting and we are done.

Otherwise, we have that $\omega$ assigns positive weight to some projections and negative weight to some other projections. Let $f$ be an operation on which $\omega'$ is defined that is not a projection, and let $k$ be the arity of $f$. If we form the superposition of $\omega$ with the sequence of functions $g_1, g_2, \ldots, g_{\text{ar}(\omega)}$, where $g_i = f$ if $\omega(e_i^{(\text{ar}(\omega))})$ is positive, and $g_i = e_1^{(k)}$ otherwise, then we obtain a new weighting $\omega[g_1, g_2, \ldots, g_{\text{ar}(\omega)}]$ which assigns positive weight to $f$ (see Example 5.3). □

Assuming that $P \neq NP$, this result tells us that *tractable* valued constraint languages are associated with certain kinds of weighted clones.

To obtain further information about the weighted clones associated with tractable valued constraint languages, we now consider some special kinds of operations. For any $k \geq 2$, a $k$-ary operation $f$ is called **sharp** if $f$ is not a projection, but the operation obtained by equating any two inputs in $f$ is a projection [20]. In other words, $f$ is sharp if for all $i, j \in \{1, \ldots, k-1\}$ with $i \neq j$, there exists an index $m \in \{1, \ldots, k-1\}$ such that $f$ satisfies the identity: $f(x_1, x_2, \ldots, x_{j-1}, x_i, x_j, x_{j+1}, \ldots, x_{k-1}) = x_m$.

**Theorem 7.5.** *Any weighted clone $W$ containing positive weightings must contain a weighting that assigns positive weight to either:*

1. *A set of unary operations that are not projections; or*

2. *A set of sharp operations.*

*Proof.* Let $\omega$ be a positive weighting in $W$ with the smallest possible arity, $k$. If $k = 1$, then we are done. Otherwise, we consider the weightings $\omega[e_1^{(k-1)}, e_2^{(k-1)}, \ldots, e_{j-1}^{(k-1)}, e_i^{(k-1)}, e_j^{(k-1)}, \ldots, e_{k-1}^{(k-1)}]$ for all $i, j \in \{1, \ldots, k\}$ with $i \neq j$.

Each of these weightings has arity $k - 1$, so, by the choice of $\omega$, must not assign positive weight to any operation except (possibly) projections. Hence all non-projection operations assigned positive weight by $\omega$ are sharp. □

We can obtain further details about these weighted clones by considering the possible types of sharp operations.

First, we observe that all sharp operations must satisfy the identity $f(x, x, \ldots, x) = x$; such operations are called **idempotent**.

Ternary sharp operations may be classified according to their values on tuples of the form $(x, x, y), (x, y, x)$ and $(y, x, x)$, which must be equal to either $x$ or $y$. There are precisely 8 possibilities, as listed in Table 1.

| Input | 1 | 2 | 3 | 4 | 5 | 6 | 7 | 8 |
|---|---|---|---|---|---|---|---|---|
| (x,x,y) | x | x | x | x | y | y | y | y |
| (x,y,x) | x | x | y | y | x | x | y | y |
| (y,x,x) | x | y | x | y | x | y | x | y |

Table 1: Sharp ternary operations

The first column in Table 1 corresponds to operations that satisfy the identities $f(x, x, y) = f(x, y, x) = f(y, x, x) = x$ for all $x, y \in D$; such operations are called **majority** operations. The last column in the table corresponds to operations that satisfy the identities $f(x, x, y) = f(x, y, x) = f(y, x, x) = y$ for all $x, y \in D$; such operations are called **minority** operations. Columns 4, 6 and 7 in Table 1 correspond to operations that satisfy the identities $f(y,y,x) = f(x,y,x) = f(y,x,x) = y$ for all $x, y \in D$ (up to permutations of inputs); such operations are called **Pixley** operations [33].

For any $k \geq 3$, a $k$-ary operation $f$ is called a **semiprojection** if it is not a projection, but there is an index $i \in \{1, \ldots, k\}$ such that $f(x_1, \ldots, x_k) = e_i^{(k)}$ for all $x_1, \ldots, x_k \in D$ such that $x_1, \ldots, x_k$ are not pairwise distinct. In other words, a semiprojection is a particular form of sharp operation where the operation obtained by equating any two inputs is always the *same* projection. Columns 2,3 and 5 in Table 1 correspond to semiprojections.

The following lemma shows that the only sharp operations of arity $k \geq 4$ are semiprojections.

**Lemma 7.6** (Świerczkowski's Lemma [36])**.** *Given an operation of arity $\geq 4$, if every operation arising from the identification of two variables is a projection, then these projections coincide.*

Hence we may refine Theorem 7.5 to obtain the following corollary.

**Corollary 7.7.** *Any weighted clone $W$ containing positive weightings must contain a weighting that assigns positive weight to either:*

1. *A set of unary operations that are not projections; or*

2. *A set of binary idempotent operations that are not projections; or*

3. *A set of ternary operations that are majority operations, minority operations, Pixley operations or semiprojections; or*

4. *A set of $k$-ary semiprojections (for some $k > 3$).*

Corollary 7.7 can be used to guide the search for tractable valued constraint languages, as we illustrate in the next section.

## 8 Classification of Boolean Valued Constraint Languages

In this section, we consider the special case of valued constraint languages over a 2-valued domain, such as the Boolean domain $D = \{0, 1\}$.

There are only four unary operations on the Boolean domain, and one of these is the projection operation $e_1^{(1)}$, which is the identity operation. The remaining three unary operations are the operations given by $f(x) = 0$, $f(x) = 1$, and $f(x) = 1 - x$. These will be referred to as constant 0, constant 1, and inversion.

There are only two binary idempotent operations on the Boolean domain that are not projections: the operations $\min$ and $\max$. The only sharp ternary operations are the unique majority operation (which we will call $\mathrm{Mjrty}$), the unique minority operation (which we will call $\mathrm{Mnrty}$), and three Pixley operations. There are no semiprojections.

Hence we can refine Corollary 7.7 even further in the special case of the Boolean domain, to limit the possibilities for weighted clones associated with tractable valued constraint languages to just nine cases.

**Theorem 8.1.** *Any weighted clone $W$ on the Boolean domain that contains positive weightings must contain a weighting $\omega$ that assigns positive weight to either:*

1. *Exactly one of the unary operations constant 0, constant 1, or inversion;*

2. *Exactly one of the binary operations $\min$ and $\max$, or both of them equally;*

3. *Exactly one of the ternary operations $\mathrm{Mjrty}$ and $\mathrm{Mnrty}$, or both of them with $\omega(\mathrm{Mjrty}) = 2\omega(\mathrm{Mnrty})$.*

*Proof.* Let $C$ be the support of $W$, and let $\omega$ be a positive weighting in $W$ with minimal possible arity. Since there are no semiprojections on the Boolean domain, Corollary 7.7 tells us that $\omega$ is either unary, binary or ternary.

Consider first the case when $\omega$ is unary. Since there are just three unary operations on the Boolean domain that are not projections, scale $\omega$ so it assigns weight -1 to the projection $e_1^{(1)}$, weight $a$ to the constant 0 operation $f_0$, weight $b$ to the constant 1 operation $f_1$, and weight $c$ to the inversion operation $f_\neg$. If $c = 1$, then $\omega$ assigns positive weight only to $f_\neg$ and we are done. Otherwise, if $f_\neg \in C$, and hence $c$ is defined, we consider the weighting $\omega' = \frac{1}{c+1}\omega + \frac{c}{c+1}\omega[f_\neg]$. It is straightforward to check that $\omega'$ assigns weight $c - 1$ to $e_1^{(1)}$, weight $a$ to $f_0$, weight $b$ to $f_1$ and weight 0 to $f_\neg$. By Lemma 6.5, $\omega'$ belongs to $W$.

If $a = 1$, then $\omega'$ assigns positive weight only to $f_0$ and we are done. Otherwise, if $f_0 \in C$, and hence $a$ is defined, we consider the weighting $\omega'' = \omega' + \frac{a}{1-a}\omega'[f_0]$. It is straightforward to check that $\omega''$ assigns positive weight only to $f_1$. By Lemma 6.5, $\omega''$ belongs to $W$.

Next consider the case when $\omega$ is binary. By Corollary 7.7 and our observations above about the possible binary idempotent operations on the Boolean domain, we know that $\omega$ assigns positive weight only to one or both of the operations $\min$ and $\max$. If either of these weights is undefined (because the corresponding function does not belong to $C$), or zero, then we are done, so assume that $\omega$ assigns positive weight to both $\min$ and $\max$. By taking the weighting $\omega + \omega[e_2^{(2)}, e_1^{(2)}]$, with a suitable scaling, we can obtain a weighting $\omega_a \in W$ that assigns weight -1 to both binary projections, weight $a$ to $\min$ and weight $2 - a$ to $\max$, for some $0 < a < 2$.

If $a < 1$, then the weighting $\omega_a + \frac{a}{1-a}\omega_a[\min, \max]$ assigns positive weight only to $\max$. If $a > 1$, then the weighting $\omega_a + \frac{2-a}{a-1}\omega_a[\min, \max]$ assigns positive weight only to $\min$. If $a = 1$, then $\omega$ assigns equal weight to $\min$ and $\max$.

Finally, we consider the case when $\omega$ is ternary. By Corollary 7.7 and our observations above about the possible ternary sharp operations on the Boolean domain, we know that $\omega$ assigns positive weight to some subset of Mjrty, Mnrty and the three Boolean Pixley operations $f_1$, $f_2$ and $f_3$ (corresponding to the fourth, sixth and seventh columns of Table 1). We note that $f_1[e_2^{(3)}, e_3^{(3)}, e_1^{(3)}] = f_3$, $f_2[e_2^{(3)}, e_3^{(3)}, e_1^{(3)}] = f_1$ and $f_3[e_2^{(3)}, e_3^{(3)}, e_1^{(3)}] = f_2$. Hence, if $\omega$ assigns positive weight to any Pixley operation, then we have that $W$ also contains the weighting $\omega' = \omega + \omega[e_2^{(3)}, e_3^{(3)}, e_1^{(3)}] + \omega[e_3^{(3)}, e_1^{(3)}, e_2^{(3)}]$ which assigns equal negative weight to each projection, and equal positive weight to each Pixley operation. By a suitable scaling we shall assume that $\omega'$ assigns weight -1 to each projection.

Suppose first that $\omega'$ assigns positive weight to at least one of Mjrty and Mnrty, and assigns weight $0 < w < 1$ to the three Pixley operations. We note that $f_i[f_1, f_2, f_3] = e_i^{(3)}$ for each $i = 1, 2, 3$. Moreover, Mjrty$[f_1, f_2, f_3] = $ Mnrty and Mnrty$[f_1, f_2, f_3] = $ Mjrty. Thus, the weighting $\omega'' = \omega' + w\omega'[f_1, f_2, f_3]$ is non-zero, assigns weight 0 to each Pixley operation and equal negative weight to all projections. By Lemma 6.5, $\omega'' \in W$.

Assume that $\omega''$ assigns positive weight to both Mnrty and Mjrty. By taking a suitable scaling, we can obtain a weighting $\omega_a \in W$ that assigns weight -1 to all three projections, weight $a$ to Mnrty and weight $3 - a$ to Mjrty, for some $0 < a < 3$.

If $a < 1$, then the weighting $\omega_a + \frac{a}{1-a}\omega_a[\text{Mjrty}, \text{Mjrty}, \text{Mnrty}]$ assigns positive weight only to Mjrty. If $a > 1$, then the weighting $\omega_a + \frac{3-a}{a-1}\omega_a[\text{Mjrty}, \text{Mjrty}, \text{Mnrty}]$ assigns positive weight only to Mnrty. If $a = 1$, then $\omega_a$ assigns positive weight to both Mjrty and Mnrty, in the ratio 2:1.

The only remaining case is when $\omega'$ assigns positive weight 1 just to the three Pixley operations. In this case we note that $f_1[e_1^{(3)}, e_2^{(3)}, f_1] = $ Mjrty, $f_2[e_1^{(3)}, e_2^{(3)}, f_1] = e_1$ and $f_3[e_1^{(3)}, e_2^{(3)}, f_1] = e_2$. Thus the function $\mu_1 = \omega[e_1, e_2, f_1]$ assigns weight $-1$ to $f_1$, $+1$ to Mjrty, and 0 otherwise. For $i = 2, 3$, we can obtain in a similar way a function $\mu_i$, which assigns weight $-1$ to $f_i$ and $+1$ to Mjrty. Then the weighting $\omega + \mu_1 + \mu_2 + \mu_3$ will assign positive weight only to Mjrty. Again, by Lemma 6.5, $\omega'' \in W$. □

Each of the nine types of weightings mentioned in Theorem 8.1 can be supported by a different clone, so these nine types of weightings can each generate different weighted clones.

Using the Galois connection developed above, this result tells us that any tractable valued constraint language over the Boolean domain must have as a weighted polymorphism one of nine specific kinds of weightings. Eight of these can be been shown to be sufficient to ensure tractability using the results of [16]. The only remaining case is the unary weighting that assigns positive weight to the inversion operation only. Our next result shows that having a weighted polymorphism of this kind is *not* a sufficient condition for tractability on its own, but if a language has any additional positive weightings as weighted polymorphisms which are not implied by this one, then it will be tractable.

**Corollary 8.2.** *Any weighted clone $W$ on the Boolean domain that contains positive weightings, satisfies exactly one of the following:*

1. $W = \mathrm{wClone}(\{\omega_\neg\} \cup \mathbf{W}_C^0)$, *for some unary weighting $\omega_\neg$ that assigns positive weight to the inversion operation only, where $C$ is the support of $W$; in this case $\mathrm{Imp}(W)$ is NP-hard.*

2. *$W$ contains one of the eight other kinds of weighting listed in Theorem 8.1; in each of these eight cases $\mathrm{Imp}(W)$ is tractable.*

*Proof.* By Theorem 8.1, $W$ must contain either a weighting $\omega_\neg$ that assigns positive weight to the inversion operation only, or at least one of the eight other kinds of weighting listed in Theorem 8.1 (or both).

If $W = \mathrm{wClone}(\{\omega_\neg\} \cup \mathbf{W}_C^0)$, for some unary weighting $\omega_\neg$ that assigns positive weight to the inversion operation only and $C$ is the support of $W$, then we are in case (1). In this case, the weighted relation $\varrho_{\neq}$ defined in Example 2.3 is an element of $\mathrm{Imp}(W)$, so $\mathrm{Imp}(W)$ is NP-hard (see the proof of Proposition 7.3).

If $W$ contains a suitable weighting $\omega_\neg$, but $W \neq \mathrm{wClone}(\{\omega_\neg\} \cup \mathbf{W}_C^0)$, then $W$ must also contain a non-zero weighting $\omega$ of minimal possible arity such that $\omega \notin \mathrm{wClone}(\{\omega_\neg\} \cup \mathbf{W}_C^0)$.

If $\omega$ is unary, then we can use the same argument as in the proof of Theorem 8.1 to show that $W$ must contain a unary operation that assigns positive weight to the constant 1 operation only or the constant 0 operation only.

If $\omega$ is not unary, then $\omega[e_1^{(1)}, \ldots, e_1^{(1)}]$ must lie in $\mathrm{wClone}(\omega_\neg)$, so $\omega$ assigns positive weights only to operations $f$ such that $f(x, \ldots, x) = x$ or $f(x, \ldots, x) = 1 - x$. If $\omega(f) = a > 0$ for some $f$ such that $f(x, \ldots, x) = 1 - x$, then we consider the weighting $\omega' = \omega + a\omega_\neg[f]$, and note that $\omega'(f) = 0$. By repeating this process we obtain a weighting $\omega''$ which assigns positive weight only to operations $f$ such that $f(x, \ldots, x) = x$. Since $\omega$ has minimal arity, these must be sharp operations, so we can proceed as in the proof of Theorem 8.1 to show that case (2) holds.

Each of the eight types of weightings in case (2) is sufficient to ensure the tractability of $\mathrm{Imp}(W)$, by the results of [16]. □

The corresponding classification for valued constraint languages over the Boolean domain was obtained in [16] using a more intricate argument involving the explicit construction of gadgets to express particular weighted relations. Here we have considered only the properties of weighted clones.

**Example 8.3.** *The weighted relation $\varrho_=$ defined in Example 2.2 has as a weighted polymorphism the weighting $\omega_{sub}$ defined in Example 5.12 which assigns equal positive weight to $\max$ and $\min$.*

*Hence the valued constraint language $\Gamma = \{\varrho_=\}$ is tractable, and remains tractable if we add to $\Gamma$ any other weighted relations that have this weighting as a weighted polymorphism. For example, we may add unary weighted relations with a single allowed value, which allow us to fix individual variables to a desired value, and still retain tractability.*

**Example 8.4.** *The weighted relation $\varrho_{\neq}$ defined in Example 2.3 has a unary weighted polymorphism that assigns positive weight only to the inversion function. It has none of the other eight types of weightings listed in Theorem 8.1.*

*It follows that $\varrho_{\neq} \notin \text{Express}(\{\varrho_{=}\})$.*

# 9 Conclusions

We have presented an algebraic theory of valued constraint languages that generalizes and extends the algebraic theory developed over the past few years to study the complexity of the classical constraint satisfaction problem. We have shown that the complexity of any valued constraint language over a finite domain with rational-valued costs is determined by certain algebraic properties which we have called weighted polymorphisms.

When the weights are all zero, the optimisation problem we are considering collapses to the classical constraint satisfaction problem, CSP. In previous work [28, 27] it has been shown that every tractable constraint language for the CSP can be characterised by an associated clone of operations. That work initiated the use of algebraic properties in the search for tractable constraint languages, an area that has seen considerable activity in recent years; see, for instance, [11, 15, 14, 13, 29, 21, 4, 2, 3, 5]. The results in this paper show that a similar result holds for the valued constraint satisfaction problem: every tractable valued constraint language is characterised by an associated weighted clone. We therefore hope that our results here will provide a similar impetus for the investigation of a much broader class of discrete optimisation problems. For example, a recent result on the power of linear programming for valued constraint languages [38] provides a characterisation of languages solvable by a standard LP relaxation in terms of weighted polymorphisms.

Many questions about the complexity of discrete optimisation problems over finite domains can now be translated into questions about the structure of weighted clones. This provides a new approach to tackling such questions by investigating the algebraic properties of weighted clones.